# EGMOF: Efficient Generation of Metal-Organic Frameworks Using a Hybrid Diffusion-Transformer Architecture

*Seunghee Han[1‡], Yeonghun Kang[2,3,4‡], Taeun Bae[1], Junho Kim[1], Younghun Kim[1] Varinia Bernales[5,6], Alan Aspuru-Guzik[2,3,5,6,7,8,9,10,11*] and Jihan Kim[1*]*

1 Department of Chemical and Biomolecular Engineering, Korea Advanced Institute of Science and Technology, Daejeon 34141, Republic of Korea.

2 Department of Chemistry, University of Toronto, 80 St. George St., Toronto, ON M5S 3H6, Canada

3 Vector Institute for Artificial Intelligence, W1140-108 College St., Schwartz Reisman Innovation Campus, Toronto, ON M5G 0C6, Canada

4 Department of Chemistry, Sungkyunkwan University, Suwon 16419, Republic of Korea

5 Acceleration Consortium, 700 University Ave., Toronto, ON M7A 2S4, Canada

6 Department of Computer Science, University of Toronto, 40 St George St., Toronto, ON M5S 2E4, Canada

7 Department of Materials Science & Engineering, University of Toronto, 184 College St., Toronto, ON M5S 3E4, Canada

8 Department of Chemical Engineering & Applied Chemistry, University of Toronto, 200 College St., Toronto, ON M5S 3E5, Canada

9 Institute of Medical Science, 1 King's College Circle, Medical Sciences Building, Room 2374, Toronto, ON M5S 1A8, Canada

10 Canadian Institute for Advanced Research (CIFAR), 661 University Ave., Toronto, ON M5G 1M1, Canada

11 NVIDIA, 431 King St. W #6th, Toronto, ON M5V 1K4, Canada

‡ These authors contributed equally

*Corresponding author: jihankim@kaist.ac.kr and alan@aspuru.com

## ABSTRACT

Designing materials with targeted properties remains challenging due to the vastness of chemical space and the scarcity of property-labeled data. While recent advances in generative models offer a promising way for inverse design, most approaches require large datasets and must be retrained for every new target property. Here, we introduce the EGMOF (Efficient Generation of MOFs), a hybrid diffusion-transformer framework that overcomes these limitations through a modular, descriptor-mediated workflow. EGMOF decomposes inverse design into two steps: (1) a one-dimensional diffusion model (Prop2Desc) that maps desired properties to chemically meaningful descriptors followed by (2) a transformer model (Desc2MOF) that generates structures from these descriptors. This modular hybrid design enables minimal retraining and maintains high accuracy even under small-data conditions. On a hydrogen uptake dataset, EGMOF achieved over 94% validity and 91% hit rate, representing significant improvements of up to 39% in validity and 29% in hit rate compared to existing methods, while remaining effective with only 1,000 training samples. Moreover, our model successfully performed conditional generation across 29 diverse property datasets, including CoREMOF, QMOF, and text-mined experimental datasets, whereas previous models have not. This work presents a data-efficient, generalizable approach to the inverse design of diverse MOFs and highlights the potential of modular inverse design workflows for broader materials discovery.

# INTRODUCTION

The potential to find needles in a haystack in the vastness of chemical space has drawn significant attention to the search for materials with desired properties[1-3]. Traditionally, new materials have been discovered through iterative and time-consuming cycles of synthesis, characterization, and testing, a process that is both costly and resource-intensive. Recent advances in artificial intelligence (AI) and specifically, machine learning (ML) have accelerated this process by enabling data-driven prediction and optimization of material properties[4, 5]. In particular, generative models have attracted growing interest as a means to directly design novel materials with targeted properties and functionalities. Various architectures, such as Generative Adversarial Networks (GANs)[6], Variational Autoencoders (VAEs)[7-9], diffusion models[10], and transformers[11], have been successfully applied to the design of new organic molecules and inorganic crystals, demonstrating their potential to revolutionize materials discovery[12-20].

However, inverse design for materials with desired properties remains challenging, as it requires vast amounts of data for effective training[21, 22]. Unlike large-scale language models like GPT and image-generation diffusion models, which are trained on billions of data points [23, 24], the amount of materials data remains scarce, and obtaining property data for these materials can be extremely expensive, both for computational simulations (e.g., density functional theory (DFT) calculations, molecular dynamics (MD) simulations)[25] as well as for experimental data[26]. Consequently, the necessary scale to allow for efficient generation of user-desired materials is difficult under these data-scarce conditions[27].

Among the several classes of materials, metal–organic frameworks (MOFs) are particularly challenging for atom-level generative modelling[28, 29]. MOFs are nanoporous materials composed of metal nodes and organic linkers, offering an enormous, chemically diverse design space[30-32]. However, their structural complexity, which consists of hundreds of atoms per unit cell, makes direct atom-level generation computationally demanding. To remedy this issue, most previous studies have adopted simplified representations, such as coarse-grained diffusion models[33], or voxel-based geometric representations[34]. More recently, building-block-aware diffusion models have been developed to assemble 3D MOFs by learning the identity and spatial arrangement of nodes and edges[35]. These approaches have two key limitations when it comes to inverse design. First, they require very large training datasets, often requiring 200,000 to 300,000 MOF structures[33-35]. This stands in contrast to the much smaller property-labeled datasets available, which include the hMOF dataset (137,652 structures)[36], CoRE MOF dataset (10,143 structures)[37], and QMOF dataset (20,373 structures)[38]. Second, these models often lack compatibility with experimental MOF datasets. These models are often restricted to hMOFs because their preprocessing pipelines demand idealized structural representations, rendering them incompatible with valuable experimental MOF datasets. The combined limitations of structural complexity, data scarcity, and incompatibility with experimental data highlight the need for a more efficient and generalized inverse design framework.

To address these challenges, we propose EGMOF (Efficient Generation of MOFs), a diffusion–transformer framework that introduces a modular inverse design approach. EGMOF overcomes the challenges of structural complexity and data scarcity by introducing a chemically informed descriptor as an intermediate representation between properties and structures. This descriptor encodes key structural and chemical features in a compact and machine-readable form, allowing efficient property-structure mapping while substantially reducing input dimensionality. Our workflow consists of two components: a diffusion model (Prop2Desc) that generates a descriptor conditioned on the target property and a transformer (Desc2MOF) that predicts the MOF structure from the generated descriptor. Because the process is modular, only Prop2Desc needs retraining when the target property changes, while the pre-trained Desc2MOF can be reused across tasks. This design dramatically reduces computational cost and training time compared to the traditional retraining of end-to-end models. Overall, EGMOF provides a data-efficient and generalizable framework for inverse design material generation by introducing a modular formulation that decouples property-conditioned representation learning from structure generation. Furthermore, by enabling flexible representation of organic building blocks, EGMOF expands the accessible chemical space.

# RESULTS

## EGMOF via Chemically-Informed Descriptors

In this work, we employ chemically informed descriptors as a low-dimensional representation of MOF structures[39, 40]. From the perspective of machine learning inputs, lower-dimensional representations are generally advantageous because they reduce the complexity of the input space and facilitate more efficient model training. Descriptors are designed to encode connectivity, local chemical environments, and 3D geometric information into compact 1D numerical vectors, enabling the use of substantially fewer input dimensions than voxel-based, graph-based, or coordinate-based representations (**Figure S1**). Beyond compactness, descriptors inherently capture chemical intuition as numerous studies have demonstrated their ability to predict diverse properties, including gas uptake, diffusivity, proton conductivity, and even text-mined quantities such as thermal or solvent removal stability[41-43]. This property-driven expressiveness highlights that descriptors retain chemically meaningful information, allowing accurate prediction across diverse material properties.

Moreover, we observe that MOFs with similar descriptors will exhibit similar properties (see Supplementary **Figure S2)**. These observations indicate that the descriptor space captures underlying structure–property relationships, enabling MOFs with different topologies or building block compositions to exhibit similar properties when their descriptor representations are similar. Accordingly, generating a descriptor corresponding to a desired property can be an efficient alternative to directly generating the full MOF structure (**Figure 1a**). Based on this insight, the Efficient Generation Model for MOFs (EGMOF) was developed, in which the descriptor serves as an intermediate representation. The model first learns to generate low-dimensional descriptors conditioned on target properties (Prop2Desc) and then predicts MOFs that match these descriptors using a pre-trained mapping module (Desc2MOF). This approach leverages two key advantages of descriptors: their low dimensionality, which reduces the complexity of the model, achieving computational efficiency and enabling effective training with limited data, and their inherent ability to capture chemical intuition allows for robust conditional generation across diverse properties.

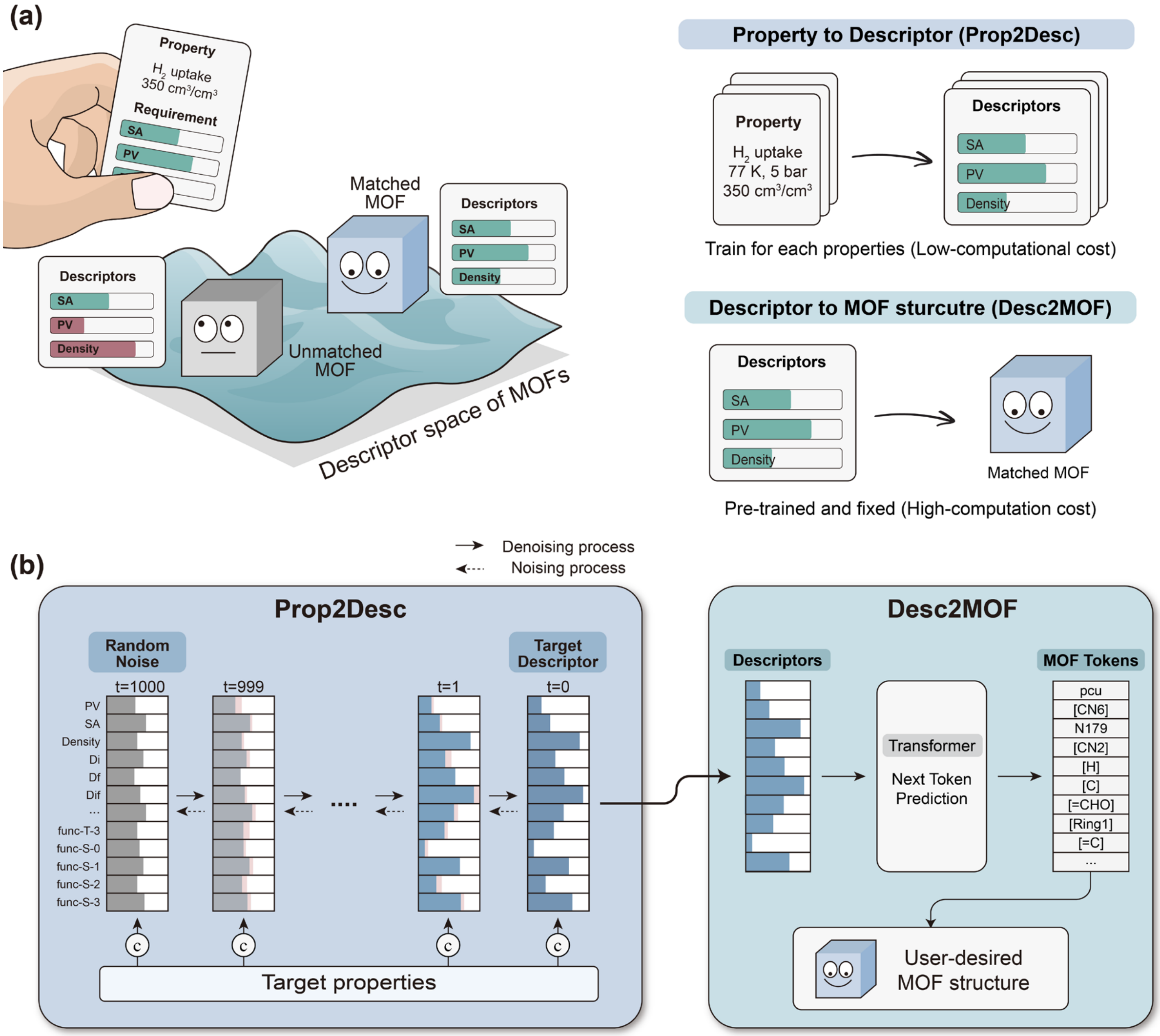

**Figure 1. Schematic Illustration of the Efficient Generation Model for MOFs (EGMOF) and its Architecture**. (a) Conceptualization of the descriptor-based inverse design process. (b) Detailed architecture showing the Prop2Desc diffusion model generating the descriptor and the Desc2MOF transformer predicting the MOF structure. The symbol "©" denotes concatenate operation.

## The Architecture and Implementation of EGMOF

The overall workflow of EGMOF is shown in **Figure 1b** and consists of two components: Prop2Desc, a diffusion model that generates chemically informed descriptors conditioned on a target property, and Desc2MOF, a transformer that reconstructs the MOF structure from the descriptors. This modular design enables efficient training and reuse: Desc2MOF is pre-trained once to learn the mapping between descriptors and MOF structures, while Prop2Desc can be re-trained independently for each new property objective. This separation allows rapid adaptation across diverse target properties without re-training the entire generative pipeline.

Prop2Desc is responsible for translating a target property into a set of chemically meaningful descriptors. Importantly, a single property does not correspond to a unique descriptor vector; instead, multiple descriptor configurations can satisfy the same property due to the inherent many-to-one relationship between structure and property. To capture this diversity, a generative modeling framework is adopted, allowing Prop2Desc to produce multiple plausible descriptor candidates for a given target property rather than a single deterministic output. To achieve this, Prop2Desc is implemented as a one-dimensional diffusion model[10] based on a U-Net[44] architecture (**Figure 1b**). Its input is a 183-dimensional descriptor vector concatenated with the target property at each diffusion timestep. During the forward process, Gaussian noise is gradually added via a Markov chain[10], while the reverse process progressively denoises the vector to recover a descriptor consistent with the specified property. Because the representation is compact and one-dimensional, Prop2Desc remains lightweight in memory and training cost while maintaining chemical interpretability. Consequently, it can generate chemically meaningful descriptors from random noise that corresponds to a given target property.

Desc2MOF takes the generated descriptors and reconstructs corresponding MOF structures, effectively translating chemical features into concrete structural representations. By pre-learning the mapping between descriptor patterns and MOF building blocks and topologies, this module significantly reduces the amount of training required during conditional generation, enabling efficient adaptation to new target properties without retraining the full model. Desc2MOF uses an encoder-decoder transformer pre-trained on 500,000 hypothetical MOFs to learn the mapping from descriptors to MOF structural tokens (**Figure 1b**)[45]. Each token represents the MOF's topology, metal node, metal edge, and organic components. In particular, organic nodes and edges are represented using SELFIES, enabling flexible and chemically valid generation of previously unseen organic building blocks. The diversity of structural components enabled by this representation is summarized in **Table S1**. These tokens can be assembled into full structures using the PORMAKE[32] Python library. The model is trained using a sequence-matching objective, achieving a token accuracy of 0.80 (**Table S2**), indicating that the mapping between descriptors and MOF components is reasonably captured. An ablation study examining the contribution of descriptor components is provided in **Table S3**. The detailed specifications of both models are provided in the Methods section.

We validated EGMOF through both unconditional and conditional generation tests for hydrogen uptake at 77 K and 5 bar (**Figure 2a)**. This condition is relevant to hydrogen working-capacity targets for onboard storage, which are commonly evaluated between 5 and 100 bar in accordance with U.S. Department of Energy (DOE) guidelines[46-49]. The Prop2Desc model was trained on 18,463 PORMAKE-generated hypothetical MOFs[45] and Desc2MOF was used to generate 1,000 structures for target volumetric uptakes ranging from 10 to 30 g/L, which were evaluated by grand canonical Monte Carlo (GCMC) simulations (details provided in the Methods section). The $H_2$ uptake dataset used for Prop2Desc training and evaluation does not overlap with the hypothetical MOF dataset used to pretrain Desc2MOF. The reliability of the GCMC simulations and associated errors is further analyzed in Supplementary **Figure S3** and **Table S4**.

The results indicate that the model can generate MOFs with properties that are consistent with the target conditions: the distribution peaks of generated MOFs align closely with their target values, corresponding to an average deviation of approximately 2.93 g/L. Unconditional generation also reproduces the training-set statistics, with a mean of 17.36 g/L and a standard deviation of 7.49 g/L, closely matching the training-set statistics (mean = 15.87 g/L and standard deviation = 7.49 g/L). This model performed best for low-range targets (10 to 15 g/L), where distributions were sharp and centered on the desired values. These quantitative results are further supported by representative structures shown in **Figure 2b**, where both unconditional and conditional generation produce MOFs whose simulated properties closely match the target values.

We further examined the model behavior at the boundaries of the training distribution. For target $H_2$ uptake values corresponding to the minimum (0 g/L) and maximum (37.77 g/L) values in the training dataset, the generated samples show distributions biased toward the target values **(Figure S4)**. While the generated samples do not always reach the target values at these boundary conditions, they exhibit distributions that are shifted toward the targets, and their corresponding descriptors occupy regions near the boundaries of the training distribution (**Figure S5**).

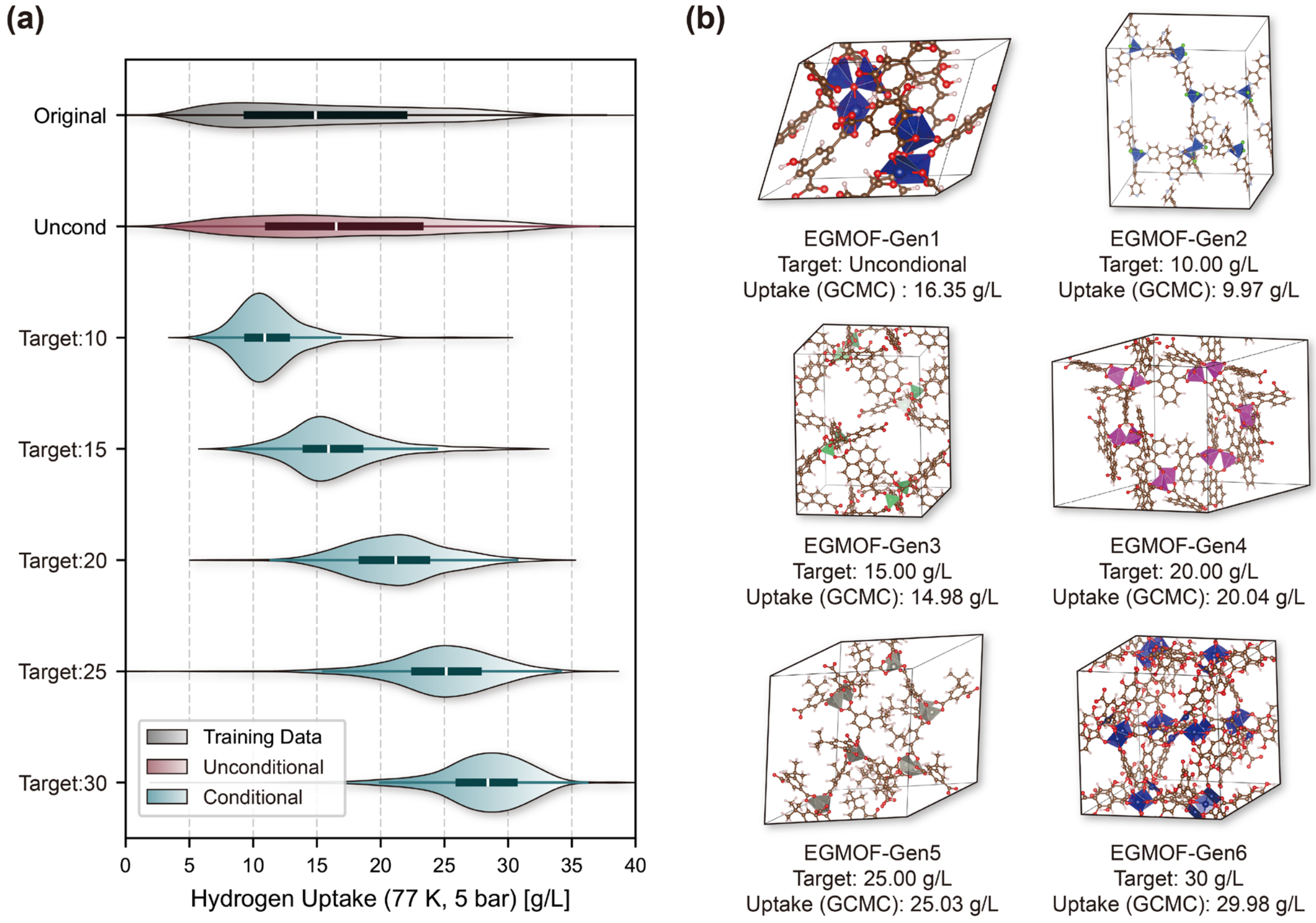

**Figure 2. Conditional Generation Results for $H_2$ Uptake at 77 K and 5 bar.** (a) Violin plots comparing $H_2$ uptake distributions for the training data, unconditional generation, and conditional generation at various target values (10 - 30 g/L). The white line within each plot represents the median, and the inner box indicates the interquartile range (IQR), spanning from the 25th to the 75th percentile. (b) Representative examples of MOF structures generated for unconditional and conditional generation. The corresponding representations of MOF tokens are provided in **Table S5**.

## Performance Comparison

We benchmarked EGMOF against state-of-the-art generative models for MOFs, including MOFDiff[33], MOFFUSION[34], and the MOFNET-based Genetic Algorithm (GA) approach[32, 50]. Only models capable of property-conditioned generation were considered (details of each model are provided in **Table S6**). For all benchmarks, the target property was hydrogen uptake at 77 K and 5 bar. To examine data efficiency, each model was trained on datasets of 1,000, 2,200, 5,000, 10,000, and 18,463 MOFs, which are orders of magnitude smaller than the 250,000 to 290,000 samples used in previous works (MOFFUSION and MOFDiff, respectively). For each target value, 1,000 MOFs were generated, and the resulting property distributions are shown in **Figures S6–**

**S9**. In addition, 10 MOF structures that pass both MOFChecker and MOSAEC evaluations containing newly introduced building blocks across different dataset sizes are illustrated in Supplementary **Figure S10**, with a summary of these structures provided in **Table S7**.

The model performance was evaluated using two metrics: validity and hit rate. Validity quantifies the proportion of structures can be assembled by PORMAKE and that successfully complete geometry optimization. Hit rate measures the proportion of generated structures whose properties fall within one standard deviation ($\pm 1\sigma$) of the target property, defined as

$$Hit\ Rate\ (\%) = P(|\hat{y} - y| \leq \varepsilon)$$

where $\hat{y}$ and $y$ denote the predicted and target property values, respectively. Here, $\varepsilon$ represents the threshold for acceptable deviation, set to one standard deviation ($\varepsilon = 1\sigma$), where σ is computed from the distribution of the target property in the training dataset. This ensures that the metric is evaluated under a consistent criterion, even when different properties are considered.

As shown in **Figure 3a and Figure 3b**, EGMOF outperforms all baselines across every dataset size. On average, EGMOF achieved the highest performance, with a validity of 94% and a hit rate of 91%, compared with the best previous models, 39% validity and 29% hit rate for the genetic algorithm. In general, other models require significantly larger datasets to achieve comparable accuracy, underscoring EGMOF's superior performance on small datasets. We further analyzed additional metrics on valid structures, including uniqueness, mean absolute error (MAE), Full Width at Half Maximum (FWHM), as well as chemical and structural validity using MOFChecker[51], MOSAEC[52], and MOFClassifier[53] (**Figure S11** and **Table S8**). The corresponding evaluation criteria are summarized in **Table S9** and **S10**, with further details provided in **Supplementary Notes S6–S8**. EGMOF consistently demonstrates strong performance across all these evaluation metrics, highlighting its ability to generate diverse, accurate, and chemically valid MOF structures.

Beyond generation quality, EGMOF demonstrates exceptional computational efficiency (**Figure 3c and 3d**). Because only the lightweight Prop2Desc module requires retraining for a new target property, the total training time was reduced by 53% and memory consumption by 82% compared to existing methods. This modular design enables rapid property-specific fine-tuning without re-training the whole model, providing a practical advantage for iterative materials discovery workflows.

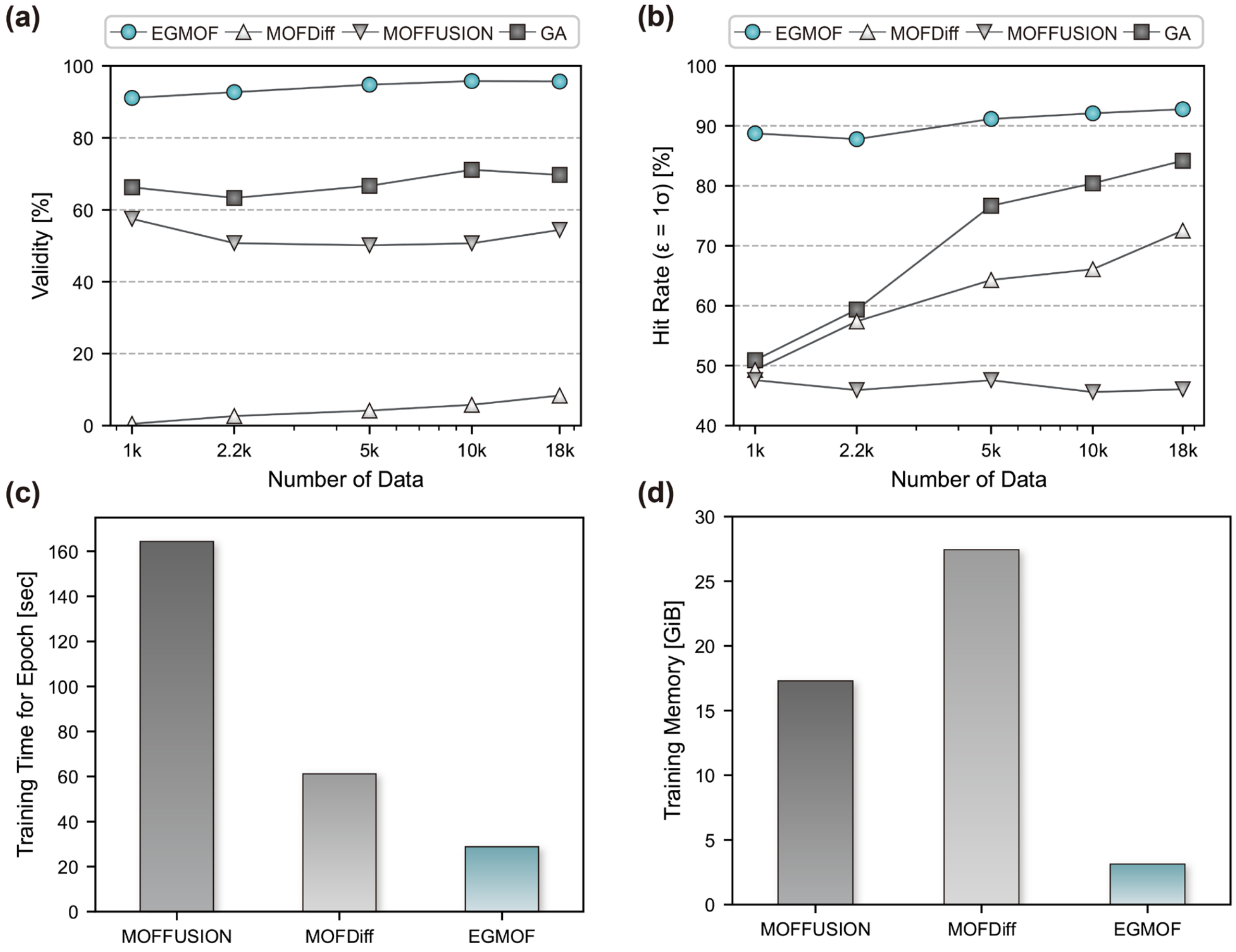

**Figure 3. Performance and Computational Efficiency Comparison of EGMOF Against Existing Generative Models (MOFDiff, MOFFUSION, and GA).** (a) Validity and (b) Hit rate ($\boldsymbol{\varepsilon} = \mathbf{1}\boldsymbol{\sigma}$) measured against varying training dataset sizes (1k - 18k). Hit ratio ($\boldsymbol{\varepsilon} = \mathbf{1}\boldsymbol{\sigma}$) is defined as the proportion of generated structures whose properties fall within one standard deviation of the target proeprty (c) Training Time per Epoch and (d) Training Memory Usage (GiB) for each model, highlighting EGMOF's superior performance and computational efficiency.

## Conditional Generation on Diverse Databases

To evaluate the generality of our framework, EGMOF was tested across 29 distinct property datasets drawn from diverse sources and prior studies based on them, including PORMAKE[45], hMOF[36], QMOF[38], CoRE[43, 54], and text-mined experimental databases[26, 41]. Further details for each dataset are provided in **Table S11**. These datasets span both computational and experimental origins and encompass a wide range of physical properties. Simulation-derived properties were evaluated from idealized crystal structures under controlled conditions, whereas experimentally reported properties were restricted to those considered less sensitive to synthesis- and measurement-dependent variability. For each dataset, the Prop2Desc module was trained to generate descriptors conditioned on three target property values: mean, mean-$\sigma$, and mean+$\sigma$, followed by Desc2MOF generation of 1,000 PORMAKE-compatible MOFs per target. Instead of computationally simulating and measuring the properties of each generated MOF, the PMTransformer was fine-tuned to estimate the properties of the generated structures, and its performance is shown in **Figure S12**[55]. To further examine the surrogate-based evaluation, additional GCMC simulations were conducted on a subset of datasets, with comparisons between PMTransformer predictions and RASPA calculations for both original and generated MOFs provided in Supplementary **Figures S13** and **S14**.

For each dataset, validity, uniqueness, hit rate, and the full width at half maximum (FWHM) of the property distributions were computed and averaged across three targets: mean, mean-$\sigma$, and mean+$\sigma$. As shown in **Figure 4a**, EGMOF consistently achieved an average hit rate of 83% across all 29 datasets, demonstrating strong robustness across properties derived from hypothetical, experimental, and text-mined sources. The corresponding distributions and results are provided in **Figure S15** and **Table S12**. In addition, EGMOF generates 120 MOF structures that pass both MOFChecker and MOSAEC evaluations containing newly introduced building blocks across these datasets, as summarized in **Table S13**.

This broad applicability highlights the ability of EGMOF to leverage experimental and literature-derived datasets for property-conditioned generation, which is an area where most existing generative models fail. Previous approaches are typically confined to hypothetical MOF datasets, since they require the decomposition of structures into discrete components (topology, nodes, edges). Such decomposition is straightforward for hypothetical MOF databases but often infeasible for experimental structures (**Table S14**). Moreover, prior conditional generation models demanded large training datasets ($10^5$ samples or more), while many experimental resources such as CoRE, QMOF, and text-mined datasets contain only $10^3$ to $10^4$, making it difficult for previous methods to utilize these existing important databases for conditioning materials generation. In contrast, EGMOF's descriptor-based modular design eliminates the need for explicit MOF decomposition during property-driven training, while retaining a building-block-based assembly scheme for structure generation, thereby enabling effective learning from small, heterogeneous datasets.

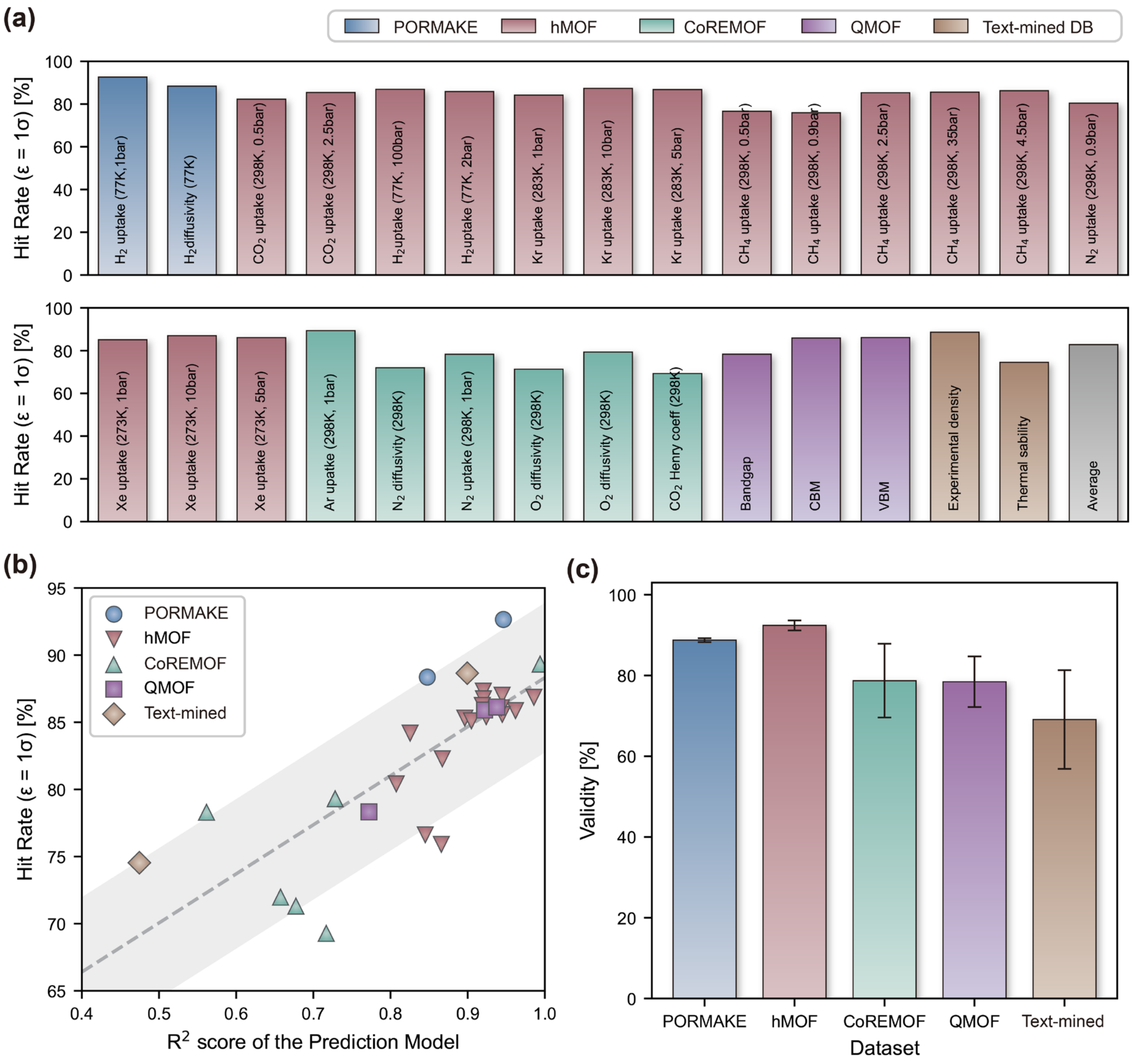

**Figure 4. Performance of the EGMOF's model Conditional Generation Across Diverse Databases.** (a) Hit Rate comparison across 29 diverse properties, showing broad applicability. Hit ratio ($\boldsymbol{\varepsilon = 1\sigma}$) is defined as the proportion of generated structures whose properties fall within one standard deviation of the target proeprty. (b) Correlation between the Hit Rate of the EGMOF generative model and $R^2$ score of the prediction model (Random Forest). The dashed line represents the trend line, and the shaded gray region indicates the area within one standard deviation of the trend. (c) Average Validity of generated structures for each source database: PORMAKE, hMOF, CoREMOF, QMOF, and Text-mined datasets. The error bars indicate the 95% confidence interval (CI) of the mean validity.

Overall, the conditional generation performance of EGMOF varied depending on the targeted property. To better understand this variation, we compared the generation performance, measured by the hit rate ($\varepsilon = 1\sigma$), with how strongly the descriptors are related to the target property. For this purpose, we used the $R^2$ score of a prediction model as a practical measure of this relationship. The $R^2$ score reflects how well a given descriptor set explains the variance of the target property[56, 57]. Although the $R^2$ score is not a direct measure of descriptor quality and can be influenced by multiple factors, it still provides a useful basis for comparison. As shown in **Figure 4b**, a higher $R^2$ score is associated with a higher hit rate for generation, indicating that performance depends on how effectively the descriptors encode information relevant to the target property. This trend suggests that when the descriptors are more strongly correlated with the target property, the model can more effectively generate candidates that satisfy the desired condition. Overall, these findings highlight the importance of chemically informative descriptors in the success of the inverse design process. Additional support is provided by the comparison of property and feature distributions between original and generated datasets across representative examples, as shown in **Figures S16–S19**.

Furthermore, EGMOF's validity was contingent on the dataset, though it maintained a high validity of over 62% across all properties. The average validity was high, exceeding 90%, for hypothetical MOFs derived from datasets such as hMOF and PORMAKE. In contrast, validity was somewhat lower for experimental datasets such as CoRE and QMOF, and for text-mined sources, likely because Desc2MOF was pre-trained exclusively on hypothetical MOFs. Previous work by Moosavi et al. showed that the descriptor spaces of experimental and hypothetical MOFs are partially disjoint, leading to minor mismatches when Desc2MOF attempts to map experimental descriptors onto nearby regions of its latent space[54].

Despite this limitation, EGMOF remains applicable to experimental datasets and is capable of performing inverse design when conditioned on experimentally relevant properties, for which many existing generative models exhibit limited applicability. For example, in the QMOF bandgap dataset (**Table S14**), existing models such as MOFFUSION and the genetic algorithm were unable to process the data due to PORMAKE representation constraints while MOFDiff discarded over 65% of the entries during preprocessing and achieved only 29% validity and 64% hit rate. In contrast, EGMOF successfully processed 85% of the data, achieving 73% validity and a hit rate of 78%. This superior performance is visually confirmed in **Figure S20**, where the MOFDiff distribution shows poor conditional generation compared to the targeted distribution generated by EGMOF.

These results confirm that EGMOF's descriptor-based design enables the effective transfer of generative capability from hypothetical to experimental spaces. This generalization is particularly valuable for properties such as bandgap, where data scarcity limits the effectiveness of traditional deep generative approaches. In addition, EGMOF generates new chemically valid organic building blocks, including 20 node types and 63 edge types, as illustrated in **Figure S21**. The synthetic complexity of these building blocks was evaluated using SCScore[58], with all values falling below 4, suggesting their potential synthetic feasibility (see **Supplementary Note S10** for details)[59]. EGMOF thus provides a practical path to the inverse design of MOFs with desired properties, even with limited experimental data.

## Descripor Analysis and Guided Decoding through Feature Importance

Analyzing the descriptors generated by the Prop2desc model provides useful insights into the chemical trends underlying target properties. This is because the generated descriptors exhibit distinct distribution patterns depending on the target property. **Figures 5a and 5b** show how the distributions of the most important descriptor for each property change as the target value varies. Void fraction represents the fraction of accessible pore volume within the structure and plays a key role in determining $H_2$ uptake. In contrast, mc-chi-0-all reflects the electronegativity of the central metal atom and strongly influences the band gap.

As the $H_2$ uptake target increases, the distribution of void fraction shifts progressively toward lower values (**Figure 5a**). In particular, for the target of 10 g/L, the distribution is concentrated around 0.7 to 0.8, whereas for 30 g/L it shifts to approximately 0.2 to 0.4, indicating that lower void fraction is favored at higher uptake targets. In contrst, the distribution of mc-chi-0-all remains largely unchanged across different target values. This trend suggests that, under 77 K and 5 bar conditions, hydrogen adsorption is influenced more strongly by energetic factors such as the overlap of adsorption potentials within the pores than by simple free volume[60]. In structures with excessively high void fraction, although the total pore volume increases, the framework density per unit volume and the degree of potential overlap decrease, which can lead to reduced volumetric uptake. Therefore, under these conditions, relatively lower void fraction and an appropriate pore environment are more favorable for achieving high $H_2$ uptake.

In contrast, for the band gap, as the target value increases, mc-chi-0-all, i.e., metal electronegativity, shows a decreasing trend (**Figure 5b**). This trend is consistent with the observation that higher electronegativity is associated with more stabilized metal centered electronic states, which can reduce the energy difference between metal and ligand based levels and lead to a smaller band gap[61]. This observation suggests that the band gap is influenced by the electronic structure difference between metal and ligand components. However, since the band gap of MOFs is determined by a combination of factors including ligand electronic structure, metal oxidation state, coordination environment, and metal ligand charge transfer, mc-chi-0-all should be interpreted as one of the descriptors reflecting these electronic structure variations rather than as a sole determining factor.

To generalize this observation, we analyzed how each descriptor influences the target property and how it contributes to the generated distribution. To quantify how much each descriptor distribution varies across different target conditions, we introduce the Earth Mover's Distance (EMD). EMD provides an intuitive measure of the difference between two distributions by calculating the minimum amount of "work" required to transform one distribution into another. A smaller EMD indicates that the distribution changes only slightly, whereas a larger EMD reflects more substantial shifts. By combining EMD with feature importance obtained from a Random Forest model, we systematically analyze how each descriptor influences the target property and how actively it is modulated during conditional generation.

**Figures 5c and 5d** show the relationship between feature importance and EMD for $H_2$ uptake and band gap, respectively. Based on these distributions, the descriptors can be categorized into three regions: Region 1, where both feature importance and EMD are high; Region 2, where feature importance is low but EMD is high; and Region 3, where EMD values are low. Although a strict correlation is not observed, descriptors with higher feature importance generally tend to exhibit larger EMD values, while less important descriptors show smaller distributional changes. This suggests that Prop2desc more actively modulates descriptors that are relevant to the target property during conditional generation.

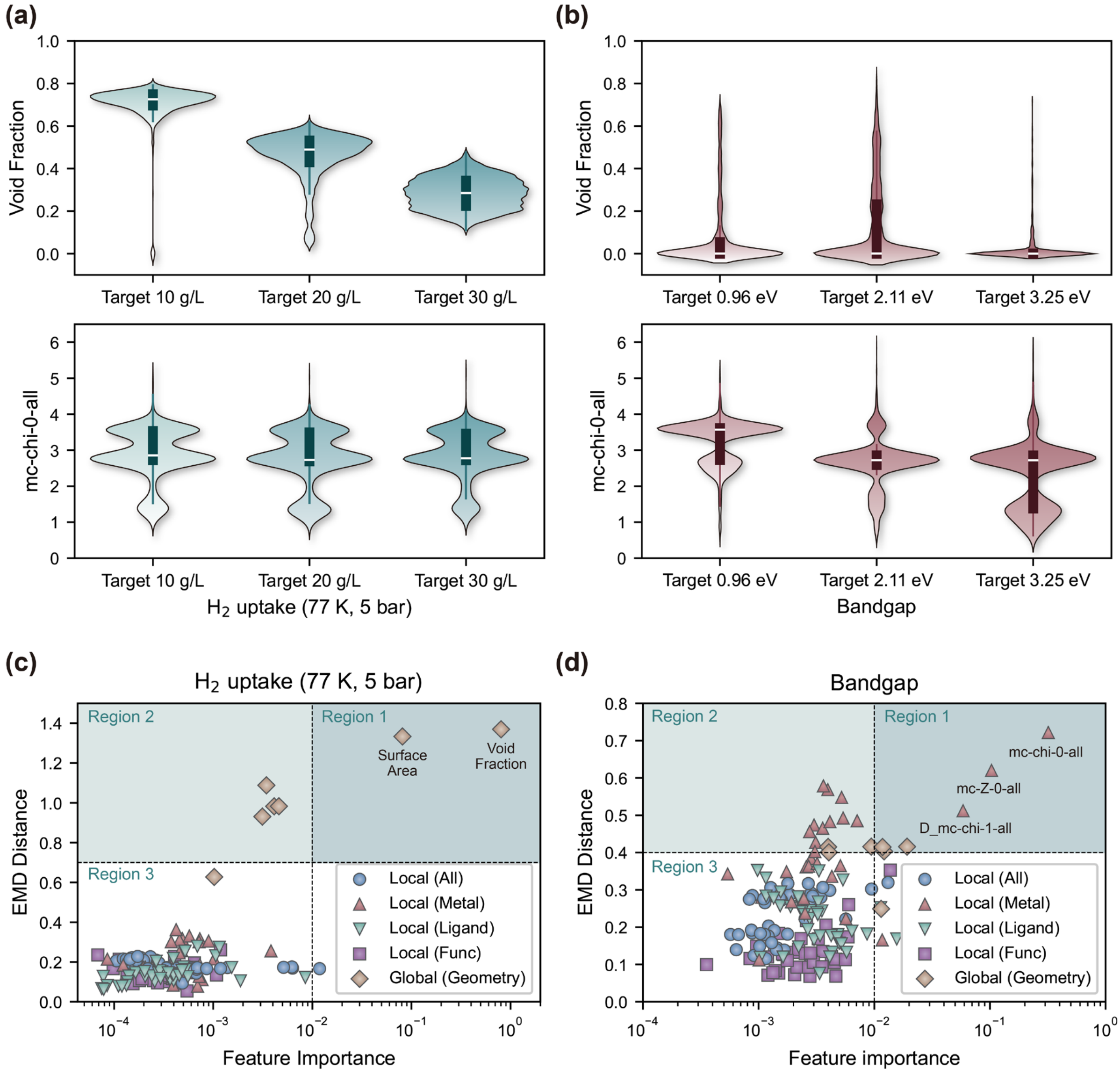

**Figure 5 Descriptor distributions and their relationship to feature importance for $H_2$ uptake and bandgap.** Distributions of void fraction and mc-chi-0-all for (a) $H_2$ uptake (10, 20, and 30 g/L) and (b) bandgap at different target values (mean-std, mean, mean+std). Feature importance versus Earth Mover's Distance (EMD) for (c) $H_2$ uptake and (d) bandgap. Region 1 denotes descriptors with both high feature importance and high EMD values, while Region 2 denotes descriptors with low feature importance but high EMD values.

Based on this relationship between feature importance and EMD, we further analyze which types of descriptors exhibit the most significant distributional shifts for each property. For $H_2$ uptake, global geometric features such as void fraction and surface area show relatively large EMD values, whereas for band gap, descriptors related to the local metal environment, such as mc-chi-0-all and related features, exhibit larger shifts. This behavior is consistent with the known dependence of $H_2$ uptake on global structural properties and band gap on local electronic environments[45]. In addition, a secondary region (Region 2) is observed, which belongs to the same feature categories as those in Region 1, with geometric features dominating $H_2$ uptake and metal local features dominating band gap. This suggests that correlated descriptors can also exhibit distributional shifts alongside primary descriptors during conditional generation. This interpretation is further supported by the Spearman correlation analysis in **Table S15**, where descriptors in Region 2 show strong correlations with those in Region 1.

Motivated by these findings, EGMOF improves the hit ratio by prioritizing descriptors that are more relevant to the target property. Instead of generating MOFs that uniformly match all descriptors in the desc2mof process, the model focuses on descriptors that are more important for achieving the desired property, thereby enhancing generation efficiency. **Figure 6a** illustrates the guided decoding strategy in Prop2desc, where feature importance is incorporated as weights to define a weighted descriptor distance, which measures the discrepancy between the predicted descriptors and the target descriptors (See Method section). During MOF token generation, the model is guided to minimize this weighted distance, steering the generation toward regions that are more relevant to the desired property.

As shown in **Figure 6b**, we compare different model variants to examine the effect of the guided decoding strategy. The unguided model already demonstrates strong performance, while introducing descriptor distance guidance further improves the hit ratio. In particular, the weighted distance strategy achieves the highest performance, resulting in a 5.44% increase in hit ratio compared to the unguided model. Notably, even the unguided EGMOF model outperforms existing baseline methods (MOFDIFF, GA, and MOFFUSION), and the guided strategies further enhance this advantage. These results demonstrate that introducing a weighted descriptor distance and guiding the generation to reduce this distance effectively improves conditional generation performance by focusing on property-relevant features.

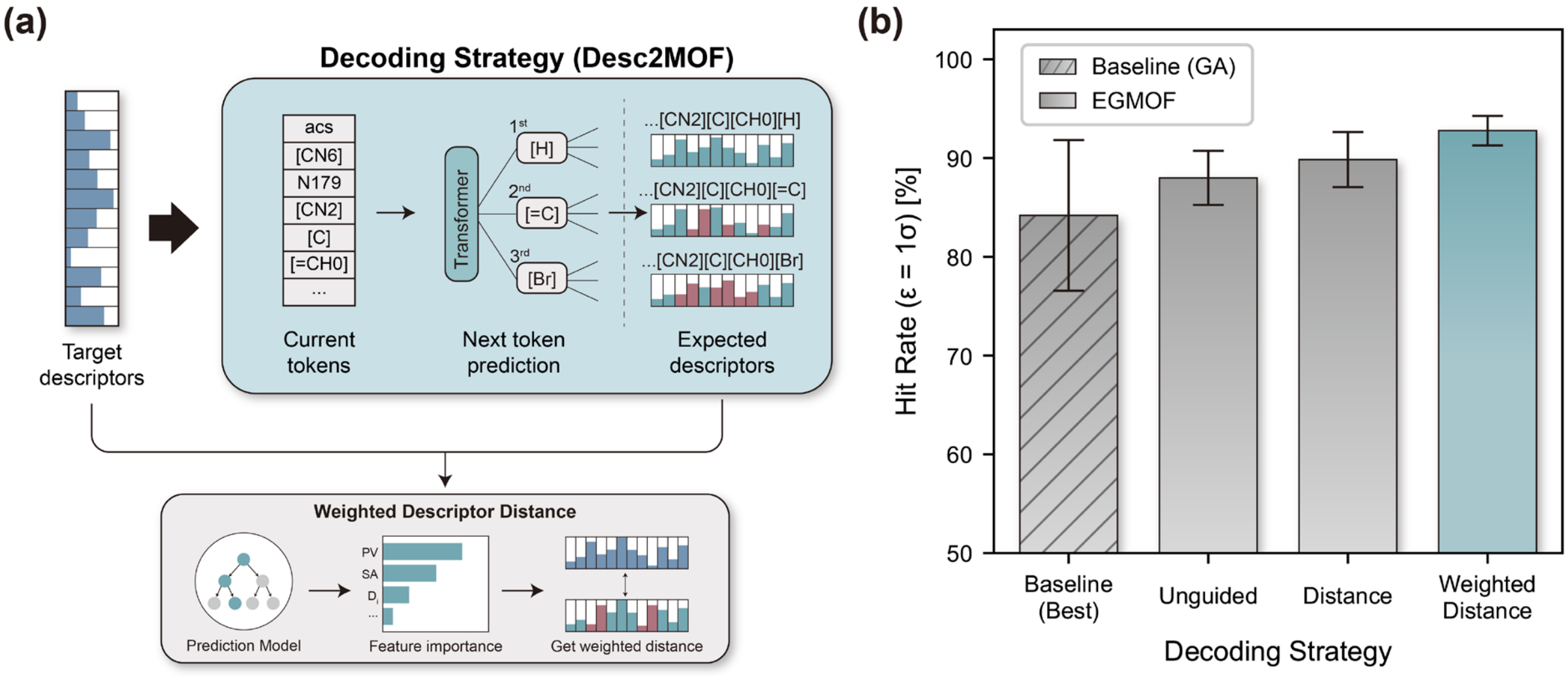

**Figure 6 Guided decoding strategy and its impact on hit ratio in EGMOF.** (a) Schematic illustration of guided decoding in Desc2MOF, where feature importance is used as weights in the weighted descriptor distance during MOF token generation. More information about the decoding strategy can be found in the Methods section. (b) Comparison of hit ratio with and without guided decoding, showing improved performance with guidance. Hit ratio ($\boldsymbol{\varepsilon = 1\sigma}$) is defined as the proportion of generated structures whose properties fall within one standard deviation of the target property.

## The Diffusion Process of Conditional Generation

To elucidate how the diffusion process navigates chemical space during conditional generation, we analyzed the denoising trajectory of descriptors, properties, and corresponding MOF characteristics of $H_2$ uptake (77 K, 5 bar) using EGMOF. As shown in **Figure 7a**, the two most important descriptors (void fraction (VF) and surface area (SA)) define the projection space, with the gray region representing the valid descriptor distribution in the training data. As denoising proceeds, the trajectory progressively moves toward the physically valid region and finally stabilizes near the subspace corresponding to the target property (e.g. 30 g/L). Such trajectories indicate that the model internalizes physically plausible pore-geometry constraints and their connection to adsorption-relevant descriptors rather than performing a simple numerical optimization. This behavior confirms that Prop2Desc refines noisy input into chemically meaningful descriptors.

Similarly, the evolution of property values during denoising (**Figure 7b**) was examined across multiple target properties (10, 15, 20, 25, and 30 g/L). After the initially erratic property predictions, the curves gradually converge toward their specified targets, demonstrating the stability and precision of the conditional guidance mechanism.

Detailed analysis of the 30 g/L target (**Figures 7c, 7d, and 7e**) illustrates the convergence dynamics more clearly. As denoising progresses, both $H_2$ uptake errors (MAE) of property prediction and the weighted descriptor distance between the generated and target descriptors steadily decrease. Concurrently, the generated MOF structure transitions from a high weighted descriptor distance, disordered state to a stable and chemically valid configuration. These results demonstrate that the diffusion process effectively guides the model from a noisy starting point to a chemically meaningful and valid structure that is consistent with the target property, and highlights EGMOF's ability to learn meaningful, property-aligned trajectories over time.

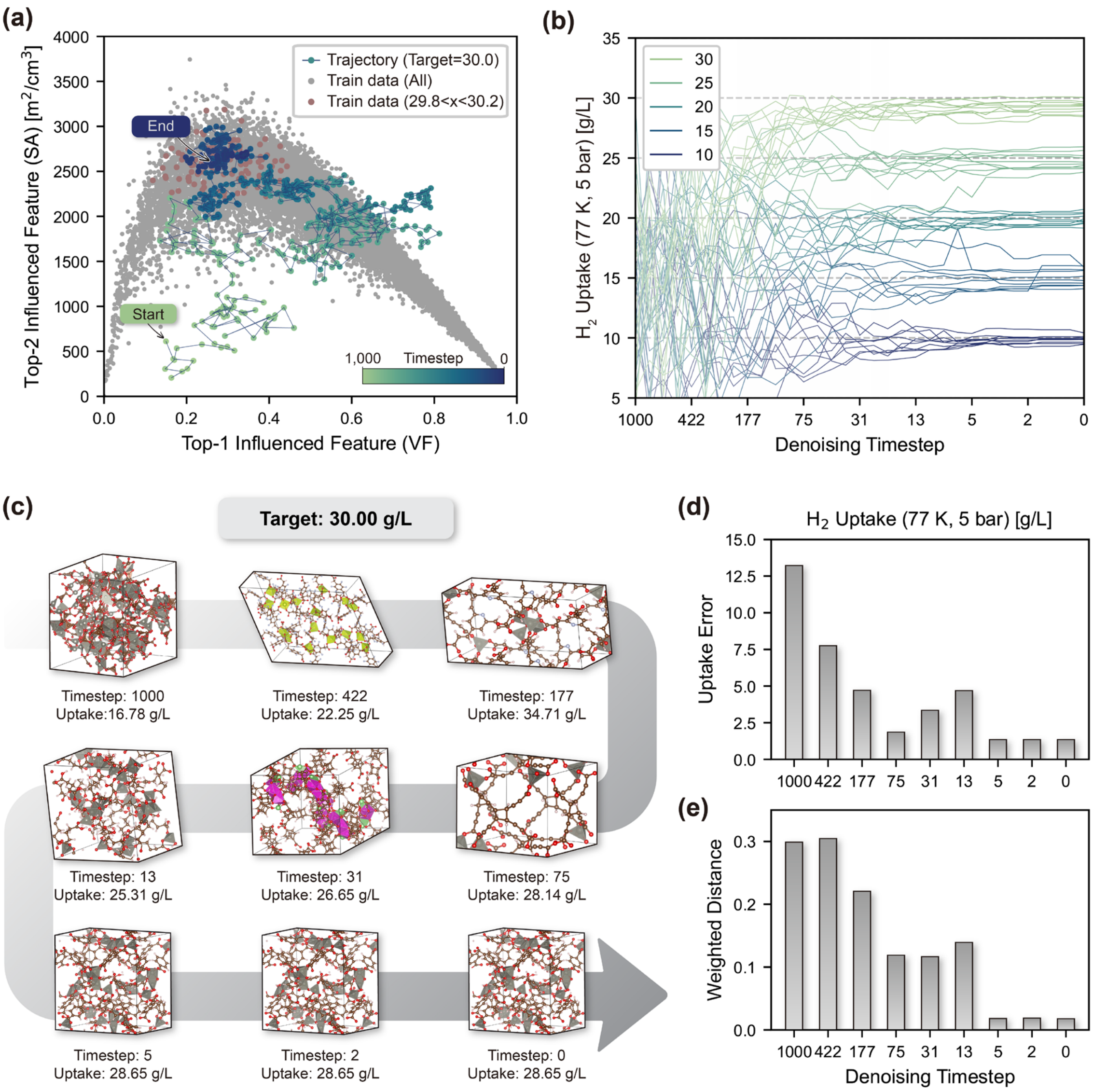

**Figure 7** Visualization of the Conditional Generation Process for $H_2$ Uptake (77 K, 5 bar). (a) Trajectory and stabilization of the MOF in the Top two descriptor VF and SA space. (b) Convergence of $H_2$ uptake values over denoising timesteps. (c) MOF structure evolution with (d) uptake error (mean absolute error, g/L) and (e) Weighted descriptor distance. Details of the weighted descriptor distance are provided in Method section. The corresponding representations of MOF Tokens are provided in **Table S16**.

# CONCLUSION

In this work, we introduce EGMOF, a data-efficient workflow that integrates Prop2Desc and Desc2MOF, using descriptors as a compact intermediary to enable efficient conditional generation. This model is applicable even to small datasets, provided the property can be represented by suitable descriptors. EGMOF achieved an average of 94% validity and 91% hit rate for an $H_2$ uptake dataset and average values of 87% validity and 83% hit rate across 29 datasets spanning both hypothetical and experimental sources. In particular, the incorporation of SELFIES enables flexible and chemically valid generation of organic building blocks, thereby expanding the accessible chemical space. EGMOF represents a substantial advancement in modular descriptor-based inverse design by effectively bridging property prediction and structure generation through interpretable descriptors. Moreover, the modular hybrid approach can be extended to other material systems that can be descriptorized, which marks an important step towards universal, data-efficient materials generation.

# METHODS

*Extract Descriptors from MOFs*

The molecular descriptors for the models were obtained by extracting revised autocorrelations (RACs) and geometric features from Crystallographic Information Files (CIFs). RACs are graph-based descriptors that capture products and differences of five heuristic atom-wise properties: nuclear charge (Z), topology (T), identity (I), covalent radius (S), and electronegativity (χ)[39]. A total of 176 RAC descriptors were generated using the MolSimplify code. In addition, seven geometric features, including void fraction (vf), cell volume (cv), density, surface area (sa), and pore size such as the largest overall diameter (di), the restricting pore diameter (df), and the largest diameter along a viable path (dif), were computed using the Zeo++ code with a probe radius of 1.2 Å[40]. Altogether, 183 descriptors were employed. Detailed information on these features is provided in **Table S17**.

*Prop2Desc*

The Prop2Desc model was developed to generate molecular descriptors conditioned on target properties by employing a diffusion-based process. The model learns a descriptor distribution through two complementary processes: a forward noising process and a reverse denoising process.

In the forward process, a clean descriptor vector $X_0 \in \mathbb{R}^{183}$ is gradually perturbed into Gaussian noise using a variance schedule $\{\beta_t\}_{t=1}^{T}$. This is formulated as a Markov chain:

$$q(X_{1:T}|X_0) = \prod_{t=1}^{T} q(X_t|X_{t-1}), \qquad q(X_t|X_{t-1}) = N(\sqrt{1-\beta_t}X_{t-1}, \beta_t I)$$

By reparameterization, the closed-form expression for directly sampling $X_t$ at any step is

$$q(X_t|X_0) = N(\sqrt{\bar{\alpha}_t}X_0, (1-\bar{\alpha}_t)I),$$

where $\bar{\alpha}_t = \prod_{s=1}^{t}(1-\beta_s)$.

The reverse process is parameterized by a neural network $\theta$ that approximates

$$p_\theta(X_{0:T}) = p(X_T)\prod_{t=1}^{T} p_\theta(X_{t-1}|X_t),$$

with the denoising distribution modeled as

$$p_\theta(X_{t-1}|X_t) = N(\mu_\theta(X_t, t), \sigma_t^2 I)$$

Training is performed by optimizing the variational bound, which simplifies to predicting the Gaussian noise added at each step. At inference, the model denoises from random Gaussian input back to the 183-dimensional descriptor space, producing descriptors consistent with the specified target properties.

The input vector is padded by one dimension, resulting in a 184 dimensional input for the model. Key hyperparameters including the learning rate, number of channels, number of U-Net layers, and the U-Net dimension reduction ratio were determined through a grid search aimed at minimizing the validation loss. The model was trained using a total of 1,000 timesteps and a batch size of 64. We employed the AdamW optimizer along with a cosine learning rate scheduler, incorporating a warm-up step of 0.05 to ensure stable initial training.

*Desc2MOF*

The Desc2MOF model was designed as a transformer-based sequence generation framework to translate continuous molecular descriptors into symbolic representations of metal–organic framework (MOF) structures. The input comprised 183 molecular descriptors (176 RAC descriptors and 7 geometric features), while the output was expressed as discrete tokens drawn from a vocabulary of 1,885 elements. This vocabulary included 1,286 topology tokens, 11 coordination number (CN) tokens, 534 metal node tokens, 20 metal edge tokens, 30 SELFIES tokens, and 4 special tokens for start-of-sequence (SOS), end-of-sequence (EOS), separator (SEP) and padding (PAD). Here, the coordination number (CN) token encodes the number of connection sites for each building block. For example, edges are always represented by [CN2], while nodes typically have CN > 2, depending on the topology. The output sequence follows a structured format in which the topology token is followed by a series of building block tokens, each paired with a CN token and separated by special separator tokens ([SEP]). Specifically, each node or edge is represented as a triplet of the form [CN#] + (N# or E# or a sequence of SELFIES tokens) + [SEP], allowing the model to encode both connectivity and chemical identity in a unified sequence. Each sequence contains up to two node tokens and two edge tokens.

The model followed an encoder–decoder architecture. The encoder consisted of a descriptor embedding layer with a hidden dimension of 256, positional encodings, and three transformer encoder layers with eight attention heads. The decoder incorporated token embeddings, positional encodings, and a three-layer transformer decoder with eight attention heads. Training was performed using the AdamW optimizer with a learning rate of 0.001, weight decay of 0.01, and a cosine learning rate scheduler with a warm-up ratio of 0.1. A batch size of 256 was employed, and the model was trained for up to 200 epochs. The loss function is a cross-entropy loss and expressed as

$$L_{CE} = -\frac{1}{N}\sum_{i=1}^{B}\sum_{t=1}^{T} 1[y_{i,t} \neq PAD] \times log p_{i,t,y_{i,t}}$$

, where $B$ is the batch size, $T$ is the sequence length, $y_{i,t}$ is the ground-truth token, and $N$ is the number of tokens. The probability distribution is given by

$$p_{i,t,v} = \frac{\exp(z_{i,t,v})}{\sum_{v'}^{V} \exp(z_{i,t,v'})}$$

with $z_{i,t} \in \mathbb{R}^V$ denoting the logits at sequence position t for sample $i$. The structural combination loss encourages valid structural decoding by penalizing probability mass assigned to invalid tokens. Denoting by $V_{i,t}$ the valid token set for position ttt under the predicted topology, this term was given as

$$L_{combi} = \frac{1}{B}\sum_{i=1}^{B}\sum_{t=1}^{T}\sum_{v \notin V_{i,t}} p_{i,t,v}$$

To enable effective conditional generation with Desc2MOF, which operates directly at inference without task-specific retraining, a large-scale pretraining stage was required. Pretraining was conducted using a generated dataset of approximately 0.5 million MOFs constructed with PORMAKE, a Python library that constructs hypothetical MOFs by combining topologies and building blocks[32]. The dataset was split into training, validation, and test subsets with a ratio of 0.70, 0.15, and 0.15, respectively.

*SELFIES Representation for Organic Building Blocks*

Organic nodes and edges are represented using SELFIES, a robust molecular string representation that guarantees syntactic validity. In this work, a vocabulary of 30 SELFIES tokens is used. During tokenization, SELFIES strings are split based on bracket units, treating each bracketed element as an individual token. To encode connectivity within organic building blocks, connection points are represented using a placeholder token [Lr], which is learned during training. The number of [Lr] tokens corresponds to the coordination number (CN) of the building block, enabling the model to capture both the number and positions of connections. Although SELFIES ensures syntactically valid strings, the generated molecules may not always correspond to the intended structures. Therefore, a validity check is performed by converting generated SELFIES to SMILES and back to SELFIES; only structures that remain invariant through this round-trip conversion are considered valid. For generated valid organic building blocks, if the resulting SELFIES corresponds to an existing building block in the PORMAKE database, it is directly reused. Otherwise, the generated structure is processed as follows: SELFIES is converted to SMILES and then to an RDKit Mol object, where the placeholder [Lr] atoms are temporarily replaced with hydrogen atoms. The structure is then geometry-optimized using the extended tight-binding (xTB) method[62]. After optimization, the hydrogen atoms are replaced back with connection points (X atoms). The finalized structure is subsequently added to the PORMAKE database for downstream assembly.

*Guided Decoding Strategy through Weighted Distance*

Conditional generation of MOFs with target properties was performed using the pretrained Desc2MOF model. Given a set of target descriptors, candidate MOF structures were autoregressively generated using beam search with a beam width of five[63]. For each input, five candidate sequences were produced, where each sequence corresponded to a tokenized representation of topology, node, and edge components.

The generated candidates were subsequently evaluated by the MOF2Desc predictor, which maps tokenized MOFs back into descriptor space (see Supporting Information for details). For each candidate, a weighted Mean-Squared Error (WMSE) between the predicted descriptors and the target descriptors was computed as

$$WMSE(x, \hat{x}) = \frac{\sum_{d=1}^{D} w_d (x_d - \hat{x}_d)^2}{\sum_{d=1}^{D} w_d}$$

where $x_d$ and $\hat{x}_d$ denote the target and predicted descriptors at dimension $d$, respectively. The weights $w_d$ were derived from feature importance values obtained from a separately trained Random Forest model that captures the relationship between descriptors and target properties. Among the five candidates, the structure with the lowest WMSE was selected as the final output for each target input. Candidates with WMSE values lower than a predefined threshold (0.5 in this work) were considered successful generations.

*Molecular simulation details*

The generated token sequences were first converted into three-dimensional MOF structures using the PORMAKE library, which assembles topology, node, and edge components into periodic frameworks. The resulting structures were then geometrically optimized using Materials Studio[64] with the Universal Force Field (UFF). Geometry optimizations were considered converged when the change in total energy was below 0.001 kcal/mol and the maximum atomic force was below 0.5 kcal/mol/Å, with a maximum of 500 optimization steps allowed.

The hydrogen uptake values used for model performance evaluation were calculated using grand canonical Monte Carlo (GCMC) simulations implemented in the RASPA package[65]. Simulations were performed at 77 K and 5 bar, employing 5,000 initialization cycles followed by 10,000 production cycles. Hydrogen molecules were treated as united atoms, and the pseudo-Feynman–Hibbs model was applied to account for quantum effects governing hydrogen behavior at low temperatures[66]. The framework atoms were described using the Universal Force Field (UFF), and cross-interactions were modeled with the Lorentz–Berthelot mixing rule[67]. A cutoff distance of 12.8 Å was employed for van der Waals interactions.

## Code availability

The code is available at **https://github.com/Yeonghun1675/EGMOF.git** and the corresponding dataset is available at **https://zenodo.org/records/19362907**

## ASSOCIATED CONTENT

**Supporting Information**. The Supporting Information is available free of charge and includes detailed descriptions of the models, datasets, and evaluation procedures used in this study. Supplementary Notes S1–S10 provide comprehensive explanations of the proposed frameworks, including MOFDIFF, MOFFUSION, the genetic algorithm, and PMTransformer, as well as supporting tools such as MOF2Desc, MOFChecker, MOSAEC, MOFClassifier and SCScore. In addition, the Supporting Information contains supplementary figures (**Figures S1–S26**) and tables (**Tables S1–S17**) that present extended analyses, model performance comparisons, validation results, and detailed data used throughout this work

## AUTHOR INFORMATION

### Corresponding Author

* Email: jihankim@kaist.ac.kr, Alan@aspuru.com

Author Contributions

S.H. and Y.K. contributed equally to this work: They conceived the research idea, designed and implemented the machine learning model architecture, and conducted the main computational experiments. T.B., J.K., and Y.K provided assistance with the implementation. A.A., V.B. and J.K. supervised the overall project. S.H. and Y.K. wrote the manuscript with editorial and discussion inputs from all co-authors. All authors have contributed to the discussions that informed the research and have given approval for the final version of the paper.

**ORCID**

Seunghee Han: 0000-0001-8696-6823

Yeonghun Kang: 0009-0001-5191-5735

Taeun Bae: 0009-0009-8275-6234

Junho Kim: 0009-0004-3141-9127

Younghun Kim: 0009-0008-8707-7401

Varinia Bernales: 0000-0002-8446-7956

Alan Aspuru-Guzik: 0000-0002-8277-4434

Jihan Kim: 0000-0002-3844-8789

**Notes**

The authors declare no competing interests.

ACKNOWLEDGEMENTS

This work was supported by the National Research Foundation of Korea (NRF) (RS-2024-00435493 and RS-2024-00451160) and by the National Supercomputing Center, which provided supercomputing resources and technical support (KSC-2024-CRE-0405). A.A.-G. thanks Anders G. Frøseth for his generous support. A.A.-G. and Y.K. acknowledge the generous support of Natural Resources Canada and the Canada 150 Research Chairs program. A.A.-G. and V.B. are supported by the University of Toronto's Acceleration Consortium, which receives funding from the CFREF-2022-00042 Canada First Research Excellence Fund. Y.K. was supported by the CIFAR AI Safety Catalyst Award (Catalyst Fund Project #CF26-AI-001). This research was enabled in part by support provided by SciNet HPC Consortium for Killaney (scinethpc.ca) and the Digital Research Alliance of Canada (alliancecan.ca).